\documentclass[twocolumn,showpacs]{revtex4}  
\usepackage{epsfig}  
  
\begin{document}  
\title{Translational Entanglement  
 of Dipole-Dipole Interacting Atoms in Optical  
Lattices }  
\author{Tom\'{a}\v{s} Opatrn\'{y}$^{1,2}$,  
Bimalendu Deb$^{3}$ and Gershon Kurizki$^{4}$}  
\address{  
$^{1}$ Department of Physics, Texas A\&M University,  
College Station, Texas 77843-4242 \\  
$^{2}$ Department of Theoretical Physics, Palack\'{y} University,  
77146 Olomouc, Czech Republic \\  
$^{3}$  Physical Research Laboratory,  
Ahmedabad-380009, India \\  
$^{4}$Department of Chemical Physics, Weizmann Institute of Science,  
76100 Rehovot, Israel   
}  
\date{\today}  
  
\begin{abstract}  
We propose and investigate a realization of the position- and   
momentum-correlated Einstein-Podolsky-Rosen (EPR) states  [Phys. Rev. {\bf 47,} 777  
(1935)] that have hitherto eluded detection. The realization involves atom  
pairs that are confined to adjacent sites of two mutually shifted optical lattices  
and are entangled via laser-induced dipole-dipole interactions.    
The EPR  
``paradox'' with translational variables is then modified by  
lattice-diffraction effects, and can be verified to a high degree of  
accuracy in this scheme.  
\end{abstract}  
  
\maketitle  
   
The ideal Einstein-Podolsky-Rosen (EPR)  \cite{EPR} state of two particles---1  
and 2, is, respectively, represented in their coordinates or momenta (in one  
dimension), as follows,  
\begin{eqnarray}  
 \langle x_1, x_2| \psi_{\rm EPR} \rangle   
 &=& \delta (x_1 - x_2), \\  
 \langle p_1, p_2| \psi_{\rm EPR} \rangle   
 &=& \delta (p_1 + p_2) .  
\end{eqnarray}  
The ``paradox'' is in the fact that  given the measured values of $x_1$ or  
$p_1$ of particle 1, one can predict the measurement result of  $x_2$   
 or $p_2$,  
respectively, with arbitrary precision, unlimited by the Heisenberg relation  
$\Delta x_2 \Delta p_2 \ge \hbar/2$. In other words, the ideal EPR state is  
fully entangled  in the continuous translational variables of the two  
particles. Approximate versions of this translational EPR state, wherein the  
$\delta$-function correlations are replaced by finite-width distributions, have  
been shown to characterize the quadratures of the two optical-field outputs of  
parametric downconversion \cite{Reid,Ou} and allow for optical  
continuous-variable teleportation \cite{BraunsteinKimble}. More recently,  
translational EPR correlations have been analyzed between dissociation  
fragments of homonuclear diatoms \cite{opa01},  whereas interacting atoms in  
Bose-Einstein condensates  have been shown to possess translational--internal  
correlations \cite{CiracZoller}.  Yet the fact remains that the original EPR  
state has eluded detection for nearly 70 years. Our goal is twofold: (i)   
propose an experimentally  feasible scheme for the creation of translational  
EPR correlations between cold atoms  that are confined in optical lattices  
\cite{Hamann} and coupled by laser-induced dipole-dipole interactions (LIDDI)  
\cite{Thirun,induceddd,duncan}; (ii) study the {\em qualitative} modifications of  
such correlations due to particle diffraction in lattices, which have been  
hitherto unexplained.   The LIDDI has been proposed as a means of  two-atom  
entanglement via their {\em internal\/} states, for quantum logic applications  
\cite{Brennen}. The ability of LIDDI to influence the spatial and momentum  
distributions of cold atoms in cavities \cite{DebKurizki}, traps and  
condensates \cite{TrapsConds}, has been investigated extensively.

To realize and measure  
the EPR translational correlations  
of material particles, one must be able to  
accomplish several challenging tasks: (a)  
switch on and off the entangling interaction;   
(b) confine their motion to single  
dimension, and (c) infer and verify the dynamical variables of   
particle 2  
{\em at the time of measurement\/} of particle 1. The latter requirement  
is particularly hard for free particles, since by the time we complete the prediction  
for particle 2, its position will have changed. In  
\cite{opa01} we suggested to   
overcome these hurdles by transforming the wavefunction of a flying   
(ionized) atom  
by an electrostatic/magnetic lens onto the image  
plane, where its  position corresponds to what it was at the   
time of the diatom dissociation.  
In this Letter we propose a different solution: (a) controlling the   
diatom formation and  
dissociation by switching on and off the LIDDI; (b)   
controlling the motion and effective masses of  
the atoms and the diatom by changing the intensities of the lattice fields.  
  
\begin{figure}[t!]  
\centerline{\epsfig{figure=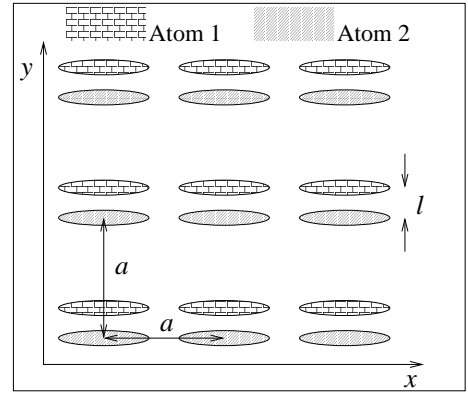,width=0.7\linewidth}}  
\caption{Proposed  
scheme of the translational EPR: two overlapping optical lattices  
displaced from each other in the $y$ direction by $l$, are sparsely  
occupied by two kinds of atoms. Each of the  
two kinds of atoms feels a different lattice; the shaded regions depict the  
energy minima (potential wells) of the lattices.   
\label{fig-lattice}  
}   
\end{figure}  
  
{\em System specification:\/}  
Let us assume two overlapping optical lattices with the same lattice constant  
$a$, as in Fig. \ref{fig-lattice}.   
The lattices are very sparsely occupied by two kinds of atoms, each kind  
interacting with only one of the two lattices.  This can be realized, e.g.,  by  
assuming two different internal (say, hyperfine) states of the atoms  
\cite{Brennen}. For both lattices, the $y$  
and $z$ directions are very strongly confining (realized by strong laser  
fields), whereas in the $x$ direction the lattice can be varied from   
moderately to weakly  
confining. Thus, the motion of each particle is confined to the  
$x$ direction. For each direction we assume that only the lowest vibrational  
energy band is occupied.   
Initially, the potential minima of the  lattices are displaced   
from each other by an amount $l \ll a$ in the $y$ direction. This enables us  
to couple the atoms of the two lattices, along $y$ using an   
auxiliary laser  
to induce the LIDDI.  
We assume the auxiliary laser to be a linearly polarized traveling wave with  wavelength $\lambda_{\rm  
C}$, moderately detuned from an atomic transition that differs from  
the one used to trap the atoms in the lattice.  The auxiliary laser propagates in the $x$  
direction and its electric field is polarized in the $y$ direction.   
The  
LIDDI potential for two identical atoms has the form \cite{Thirun}  
\begin{eqnarray}  
V_{\rm dd}=-V_{\rm C} F_{\theta}(k R),\ \quad  
 F_{\theta}(k R) = \cos \left(   
 k R \cos \theta \right) \nonumber \\  
 \times  
 \left\{ (2\! -\! 3\cos^2 \theta) \left[ \frac{\cos k R}  
 {(k R)^3}\!  + \! \frac{\sin k R}{(k R)^2} \right]\!   
 + \! \cos^2 \theta \frac{\cos k R}{k R}  
 \right\}.  
\end{eqnarray}  
Here $V_{\rm C} = \alpha^2  
k^3 I_{\rm C}/(4 \pi \epsilon_0^2 c)$,  
where $k=2\pi/\lambda_{\rm C}$, $I_{\rm C}$  is   
the coupling laser intensity,  and the atomic dynamic  
polarizability  is 
$\alpha = 2 \omega_A |\mu|^2/[\hbar (\omega_A^2-\omega^2)]$,  
$\mu$ being the  
dipole moment element, $\omega_A$ the atomic transition frequency, and  
$\omega = kc$. The position-dependent part $F_{\theta}(k R)$ is a  
function of $R$, the distance between the atoms, and $\theta$, the angle between the interatomic axis and   
the wavevector of  
the coupling laser.   
Since $l\ll 2a$, $V_{\rm dd}(R)$  
 has a pronounced minimum for atoms located at  
the nearest sites, $R\simeq l$, where $V_{\rm dd}(R)\simeq  
-V_{\rm C}(\lambda_{\rm C}/l)^3 /(4 \pi^3)$.  
Under these assumptions, we can treat the system as consisting   
of pairs of ``tubes'', that are oriented along $x$, either  
empty or occupied.  
Only atoms within adjacent tubes are appreciably attracted to each  
other  
along $y$, due to the LIDDI.  
  
{\em EPR states:}  
We now focus on the subensemble of tube-pairs in which each tube  is  
occupied by exactly one atom.   
In the absence of LIDDI,  
the state of each atom can be described in terms of the  
Wannier functions $|\chi_j\rangle$ \cite{Wannier} that are localized   
at lattice sites with index $j$ and may hop  
to the neighboring site at the rate $V_{\rm hop}/\hbar$, where   
$V_{\rm hop}=\langle \chi_j|\hat H_{\rm  
lat}|\chi_{j+1}\rangle$, and $\hat H_{\rm lat}$ is the lattice   
Hamiltonian.  
In a 1D lattice,  
$\hat H_{\rm lat} = (U_0/2) \cos \left( 2\pi x/a \right) + \hat p_x^2/(2m)$,  $m$  
being the atomic mass.    
The hopping rate is  
related to the energy bandwidth of the lowest lattice band $V_{B}$ by $V_{B}\approx 4  
|V_{\rm hop}|$ (for exact expressions see \cite{Slater}). For a shallow lattice potential ($U_0 \lesssim 15 E_{\rm rec}$) we  
may use the approximate formula $V_{\rm hop} \approx E_{\rm rec} \exp (- 0.26\  
U_0/E_{\rm rec})$, where the recoil energy is  $E_{\rm rec} = 2\pi^2 \hbar^2/(m  
\lambda_{\rm L}^2)$.   
  
Let us switch on the LIDDI, so that $|V_{\rm hop}| \ll |V_{\rm  
dd}|$.  
Then the ground state of such a tightly bound diatom can be  approximated by    
$ |\psi_0\rangle \propto  
 \sum_{j}|\chi_{j}^{(1)}\rangle|\chi_{j}^{(2)}\rangle$.   
This means that when particle 1 is found at the   
$j$th site of lattice 1,   
then particle 2 is found  
at the  
$j$th site of lattice 2, with  
position dispersion given by the half-width $\sigma$ of the atomic  
(Gaussian-like) Wannier  
function in the lowest band, $\sigma^2 = \hbar  
\lambda_{\rm L}/(4\pi\sqrt{mU_0})$.   
To next order in $V_{\rm hop}/V_{\rm dd}$, the nonzero probability  
of atoms to be located at more distant sites changes the diatomic   
position (separation) dispersion   
to  
$ \Delta x_{-}^2 \approx \sigma^2 + 2 a^2 \left(   
 \frac{V_{\rm hop}}{V_{\rm dd}} \right) ^2$.

The  
states of the tightly bound diatom form a separate band whose  
bandwidth is $V_{B}^{\rm (2 at)} \approx 4 |V_{\rm  
hop}^{\rm (2 at)}|$,  
below the lowest atomic  
vibrational band.   
The {\em diatomic\/}   
hopping potential $V_{\rm hop}^{\rm (2  
at)}$ can be found by   
assuming that the two atoms  
consecutively  hop to their neighboring sites, i.e., the state change   
$|\chi_{j}^{(1)}\rangle|\chi_{j}^{(2)}\rangle \to   
|\chi_{j+1}^{(1)}\rangle|\chi_{j+1}^{(2)}\rangle$ is realized   
either via  
$|\chi_{j}^{(1)}\rangle|\chi_{j}^{(2)}\rangle \to   
|\chi_{j+1}^{(1)}\rangle|\chi_{j}^{(2)}\rangle \to   
|\chi_{j+1}^{(1)}\rangle|\chi_{j+1}^{(2)}\rangle$, or via  
$|\chi_{j}^{(1)}\rangle|\chi_{j}^{(2)}\rangle \to   
|\chi_{j}^{(1)}\rangle|\chi_{j+1}^{(2)}\rangle \to   
|\chi_{j+1}^{(1)}\rangle|\chi_{j+1}^{(2)}\rangle$.   
By adiabatic elimination of the higher-energy intermediate states  
one obtains $V_{\rm  
hop}^{\rm (2 at)} \approx 2 V_{\rm hop}^2 /V_{\rm dd}$.   
  
To realize a momentum  
anti-correlated EPR state, the temperature of the system must satisfy   
$k_B T \ll V_{B}^{\rm (2 at)}$.  
The dependence of the momentum anti-correlation on temperature   
is given by $\Delta p_{+}^2/  
(2 m_{\rm eff}^{\rm (2 at)}) \approx \frac{1}{2} k_{B} T$,   
where we have introduced the sum-momentum spread $\Delta p_{+}$ and  
(analogously to atomic effective mass \cite{Slater})  
the two-atom effective mass  
$ m_{\rm eff}^{\rm (2 at)} = \frac{2\hbar^2}{V_{B}^{\rm (2 at)}a^2}  
 \approx \frac{\hbar^2 |V_{\rm dd}|}{4V_{\rm hop}^2 a^2}$.  
We then obtain  
$ \Delta p_{+}^2 \approx \frac{\hbar^2 |V_{\rm dd}|}{4V_{\rm hop}^2 a^2}  
 k_B T$.  
  
\begin{figure}[t!]  
\centerline{\epsfig{figure=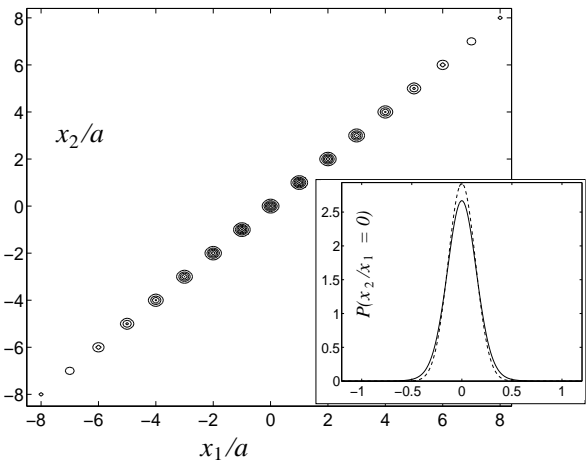,width=0.9\linewidth}}  
\caption{  
Joint probability distribution of the positions of two lithium atoms   
in adjacent  
optical lattices, prepared in a diatom state as specified in the   
text, using  
the ground state of the external harmonic potential with half-width of $\sigma_E=6a$  
and temperature of 10~nK.  
Inset:  
Position probability of atom 2 in the state above, conditional on   
atom 1  
being measured at site 0 (full line).   
Dashed line: Gaussian approximation of the  
Wannier function with the half-width $\sigma = 0.14 a$.   
\label{figconprobx}  
}   
\end{figure}  
  
\begin{figure}[t!]  
\centerline{\epsfig{figure=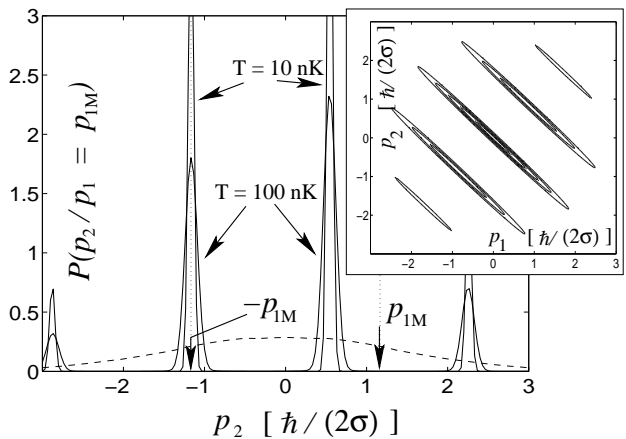,width=0.95\linewidth}}  
\caption{  
Conditional probability of the momentum of atom 2 after the momentum of atom 1 has been  
measured for lithium diatoms prepared as in the text (the measured value $p_1=p_{1{\rm M}}$ is indicated with an arrow).    
The dashed line  
corresponds to the marginal probability distribution of momentum $p_2$  
irrespective of the momentum of atom 1 at the temperature $T=100$ nK.  
The half-width of each peak is equal  
to $1/s$ of Eq. (\ref{sparameter}).    
Inset: joint  
probability distribution of the atomic momenta in the state above with  $T=100$~nK.  
\label{figconprobp}  
}   
\end{figure}

Although the values $\Delta x_-^2$ and $\Delta p_{+}^2$ estimated above   
refer to the respective peak  
widths, they principally differ from the position and momentum   
uncertainties of free particles:  
due to the lattice periodicity, the {\em position and momentum  
distributions have generally a multi-peak structure}. The two-particle  
joint position distribution of the ground state is   
a chain of peaks of  
half-width $\sigma$ separated by $a$; the peaks are located along the line  
$x_2=x_1$ (Fig. \ref{figconprobx}). The corresponding joint momentum distribution spreads over an area of  
half-width $\hbar/(2\sigma)$ and consists of ridges in the direction $p_2=-p_1$.  
These ridges are separated by $2\pi \hbar/a$, and for a lattice of $N$ sites,  
the half-width of each ridge is $\pi \hbar/(Na)$ (Fig.   
\ref{figconprobp}).

To evaluate how ``strong'' the EPR effect is, we compare the product of  
the half-widths of the position and momentum peaks in the tightly  
bound diatom state  
described above  
with the limit of the Heisenberg uncertainty relations, $\Delta x  
\Delta p \ge \hbar/2$, defining the parameter $s$ \cite{opa01}:  
\begin{eqnarray}   
s = \frac{\hbar}{2\Delta x_{-}\Delta 
 p_{+}}. 
 \label{sparameter}   
\end{eqnarray}   
A value of $s$ higher than 1 indicates the occurrence of the EPR effect; the  
higher the value of $s$, the stronger the effect. Strictly speaking, because of  
the multi-peak momentum distribution, one should not use the original  
form of Heisenberg uncertainty relations but  a more general relation, as  
discussed, e.g., in \cite{Uffink}, that distinguishes the  
uncertainty of  
a few narrow peaks from that of a single broad peak.  
However, even the simple  
half-width of the peaks  
is a useful measure of the  
EPR effect.  
In order to maximize $s$, we must adhere to the  
trade-off between making $\Delta x_{-}$ as small as possible, by decreasing  
$|V_{\rm hop}/V_{\rm dd}|$, and making $\Delta p_{+}$ as small as possible, by  
increasing $|V_{\rm hop}/V_{\rm dd}|$. The optimum value of $s$ generally   
depends on  
the lowest available temperature of the diatom, as detailed below.  
  
{\em EPR state preparation:} Cooling down the diatomic system  to prepare the  
EPR state is a non-trivial task. We suggest to attack the problem by   
a three-step  
approach. (i) Let us switch off both the LIDDI   
and the $x$-lattices, switch on an external,  
shallow, harmonic potential in the $x$ direction, and cool the   
$x$-motion  
of the atoms down to the ground state of the external  
potential. The width $\sigma_E$ of the ground state should be several times the  
lattice constant; it is related to the desired momentum anti-correlation by  
$\sigma_E \approx \hbar/(\sqrt{2} \Delta p_{+})$.  The temperature must be  
$T\ll \hbar^2/(4m k_B \sigma_E^2)$. (ii)  
A weak lattice potential in the $x$-direction is then slowly switched   
on, so that the state becomes  $\approx (\sum_{j}\alpha_j  
|\chi_{j}^{(1)}\rangle ) (\sum_{l}\alpha_l|\chi_{l}^{(2)}\rangle ) =   
\sum_{j}\alpha_j^2 |\chi_{j}^{(1)}\rangle |\chi_{j}^{(2)}\rangle + \sum_{j\neq  
l}\alpha_j \alpha_l |\chi_{j}^{(1)}\rangle  |\chi_{l}^{(2)}\rangle$, where the  
coefficients $\alpha_j \sim \exp [-(j-j_0)^2 a^2/(4 \sigma_E^2)]$ are  
Gaussians localized  around the minimum of the external potential.   
(iii) We  
switch on the LIDDI and change the sign of the external potential,  
from attractive to   
repulsive, acting to remove the particles from the lattice.  The  
two parts of the wavefunction would behave in different ways. The    
paired atoms, corresponding to the part of the wavefunction   
$\sum_{j}\alpha_j^2 |\chi_{j}^{(1)}\rangle |\chi_{j}^{(2)}\rangle$,  
move slowly because of their large effective mass $m_{\rm eff}^{\rm (2 at)}$,  
whereas single (unpaired) atoms, because of their smaller effective   
mass, $m_{\rm eff} \ll m_{\rm  
eff}^{\rm (2 at)}= |V_{\rm dd}|/ (2V_{\rm hop})m_{\rm eff}$, are   
ejected out of  
the lattice  and separated from the diatoms as glumes from   
grains.    
The paired atoms remaining in the lattice are then in the state  
$\sim \exp [-(j-j_0)^2 a^2/(2 \sigma_0^2)]  
|\chi_{j}^{(1)}\rangle |\chi_{j}^{(2)}\rangle$ wherein positions are correlated  
with uncertainty $\Delta x_+\approx \sigma_E/\sqrt{2}$ and momentum uncertainty   
$\Delta p_{+}\approx\hbar/\Delta x_+$.  
At higher temperatures the atoms are not cooled to the ground state of the  
external potential and the momentum anti-correlation becomes $\Delta p_{+}  
\approx\hbar/\{ \sqrt{2}\sigma_E  \tanh [\hbar^2/(2\sigma_E^2 mk_B T)]\}$. The  
parameter $s$ of Eq. (\ref{sparameter}) can then be estimated as  
\begin{eqnarray}  
 s \approx \frac{\sigma_E}{\sqrt{2} \sigma}  
 \tanh \left[ \frac{1}{\pi^2} \left( \frac{a}{\sigma_E} \right)  
 ^2 \frac{E_{\rm rec}}{k_B T} .  
\right]  
\end{eqnarray}  
This equation enables us to select the optimum external harmonic potential  
 (specified here by $\sigma_E$)   
 such that the parameter $s$ is maximized,  
 under the constraint of the lowest achievable  
 temperature $T$.  
  
The small effective mass of unpaired atoms allows us to cool them  
individually, restricting their cooling to temperatures higher than   
that corresponding to the bottom of the diatomic band. The  
price is, however, that most of the atoms are discarded and only   
a small  
fraction remains in the diatom state. Specifically,  out of the total  
number of tube pairs occupied by two atoms, a fraction of $\sim  
a/\sigma_E$ will remain in the bound diatom state.  
The  
different behavior of the paired vs. unpaired atoms in a periodic   
potential is a sparse-lattice analogy of the Mott-insulator vs. superfluid  
state of the fully occupied lattice, recently observed in Ref. \cite{Greiner}.

{\em Measurements:\/}  
After preparing the system in the EPR state, one can test its properties  
experimentally. To this end we may increase the lattice potential $U_{0}$,  
switch off the field inducing the LIDDI, and separate the two lattices by  
changing the laser-beam angles. By increasing $U_0$, the atoms lose their  
hopping ability and their quantum state is  
``frozen'' with a large  effective mass: the  
bandwidth $V_B$ decreases exponentially with $U_0$ and the effective   
mass increases exponentially,  
so that the atoms become too ``heavy'' to move. One has then   
enough time to perform measurements on each of them.   
  
The atomic position can be measured by detecting its resonance   
fluorescence.  
After finding the site occupied by atom 1, one can  
infer the position of atom 2. If this inference is confirmed in a large  
ensemble of measurements, it would suggest that there is an ``element of  
reality'' \cite{EPR} corresponding to the position of particle 2.    
The atomic  
momentum can be measured by switching off the   
$x$-lattice potential of the  
measured atom (thus bringing it back to its ``normal'' mass $m$): the distance  
traversed by the atom during a fixed time is proportional to its  
momentum.   
One can test the EPR correlations between the  
atomic distributions occupying the two lattices: a large number of pairs  
would be tested in a single run.  
The correlations in $x$ and anti-correlations in $p$ would be  
observed by matching the distribution histograms measured on  
particles from the two lattices.

{\em Example:\/} We consider two lithium atoms in two lattices with 
$\lambda_{\rm L}=$ 323 nm (corresponding to the transition 2s--3p) and a 
dipole-dipole coupling field of $\lambda_{\rm C}=$ 670.8 nm (transition 
2s--2p). The field intensities are $I_{\rm L} =$ 0.35 W/cm$^2$ and $I_{\rm 
C} =$ 0.1 W/cm$^2$, and the field detunings are $\delta_{\rm L} = 50 
\gamma_{\rm L}$, $\delta_{\rm C} = 100 \gamma_{\rm C}$, the decay rates 
being $\gamma_{\rm L} = 1.2 \times 10^6$ s$^{-1}$, and $\gamma_{\rm C} = 
3.7 \times 10^7$ s$^{-1}$. The two lattices are displaced by $l=$ 40~nm. 
{F}rom these values we get the lattice potential $U_0 = 7.42 E_{\rm rec}$, 
the dipole-dipole potential of the nearest atoms $V_{\rm dd} = -2.16 
E_{\rm rec}$, and the hopping potential $V_{\rm hop}=-0.0355 E_{\rm rec}$. 
The two-particle hopping potential is then $V_{\rm hop}^{\rm (2at)}\approx 
-0.0012 E_{\rm rec}$ and the ratio of effective masses of a diatom and a 
of single atom is $m_{\rm eff}^{\rm (2at)} /m_{\rm eff} \approx 30$. The 
position uncertainty of atom 2, after position measurement of atom 1, is 
then $\Delta x_{-} \approx \sigma = 0.136 a = 22$~nm (see Fig.    
\ref{figconprobx}).  The correlated pairs are prepared by first cooling 
independent atoms in an external harmonic potential with the ground-state 
half-width of $\sigma_E = 6a$ (frequency of 1.2 kHz $\sim 33$~nK).  After 
the unpaired atoms are removed from the lattice, we calculate the momentum 
distribution for two different temperatures, 10 nK and 100 nK. The 
conditional probability of momentum $p_2$ of particle 2, provided that the 
momentum of particle 1 was measured as $p_{1M}$ is plotted in Fig. 
\ref{figconprobp}. The resulting half-widths of the peaks can be used to 
find the parameter $s$; we have $s \approx 30$ for $T=$ 10~nK, and 
$s\approx 11$ for $T=$ 100~nK. Note that in current optical experiments 
\cite{Ou} $s\lesssim 4$.

To sum up, the proposed scheme is based on the adaptation of existing  
techniques (optical trapping, cooling, controlled dipole-dipole  
interaction) to the needs of atom-atom translational entanglement.   
The most important feature of the scheme  
is the manipulation of the effective mass, both for the EPR-pair   
preparation (by separating the  
``light'' unpaired atoms from the ``heavy'' diatoms)  
and for their detection (by ``freezing'' the atoms  
in their initial state so that their EPR correlations  
are preserved long enough).  
This scheme has the capacity of demonstrating the original  
EPR effect for positions and momenta, as discussed in the classic paper  
\cite{EPR}.  
A novel element of the present scheme is the extension of the EPR  
correlations to   
account for lattice-diffraction effects.  
Applications of this approach to matter teleportation \cite{opa01}  
and quantum   
computation with continuous  
variables \cite{comput} can be envisioned.  
The fact that our system represents a blend of continuous and  
discrete variables may be utilized for quantum 
information-processing (to be discussed elsewhere).

{\bf Acknowledgments:} We acknowledge the support of the US-Israel   
BSF, Minerva and the EU Networks QUACS and ATESIT.



\begin{references}  
  
\bibitem{EPR}  
A. Einstein, B. Podolsky, and N. Rosen,  
Phys. Rev. {\bf 47,} 777 (1935).  
  
\bibitem{Reid}  
M.D. Reid and P.D. Drummond,   
Phys. Rev. Lett. {\bf 60,} 2731 (1988).  
  
\bibitem{Ou}  
Z.Y. Ou, S.F. Pereira, H.J. Kimble, and K. C. Peng,  
Phys. Rev. Lett. {\bf 68,} 3663 (1992).   
  
\bibitem{BraunsteinKimble}  
S.L. Braunstein and H.J. Kimble,  
Phys. Rev. Lett. {\bf 80,} 869 (1998).  
  
  
\bibitem{opa01}  
T. Opatrn\'{y} and G. Kurizki, Phys. Rev. Lett. {\bf 86}, 3180 (2001).  
  
\bibitem{CiracZoller}  
L.-M. Duan  
et al.,  
Phys. Rev. Lett. {\bf 85,} 3991 (2000); H. Pu and   
P. Meystre, ibid {\bf 85,} 3987 (2000).  
  
\bibitem{Hamann}  
S.E. Hamann et al., Phys. Rev. Lett. {\bf 80,} 4149 (1998);  
A. Fioretti et al., Phys. Rev. Lett. {\bf 80,} 4402 (1998).  
  
\bibitem{Thirun}  
T. Thirunamachandran, Molecular Physics {\bf 40}, 393 (1980).  
  
\bibitem{induceddd}   
M.M. Burns  
et al., Science {\bf 249}, 749 (1990);  
P.W. Milonni and A. Smith, Phys.Rev. A {\bf 53},  
3484 (1996).  
  
  
\bibitem{duncan} D. O'Dell  
et al.,  
Phys.Rev.Lett. {\bf 84}, 5687 (2000);  
S. Giovanazzi et al.,  
Phys. Rev. A {\bf 63}, 031603(R) (2001).  
  
  
\bibitem{Brennen} G.K. Brennen et al.,  
Phys. Rev. Lett. {\bf 82,} 1060 (1999);  
I.H. Deutsch and G.K. Brennen Fortschr. Phys. {\bf 48,} 925 (2000);  
O. Mandel et al., cond-mat/0301169 (2003).  
  
  
\bibitem{DebKurizki}  
B. Deb and G. Kurizki, Phys. Rev. Lett. {\bf 83,} 714 (1999).  
  
\bibitem{TrapsConds}  
G. Kurizki et al., Lecture Notes in Physics,  
vol. {\bf 601,} p. 388 (Springer, 2002);  
S. Giovanazzi, D. O'Dell, and G. Kurizki,   
Phys. Rev. Lett. {\bf 88,} 130402 (2002).   
  
  
\bibitem{Slater}  
J.C. Slater, Phys. Rev. {\bf 87,} 807 (1952).  
  
\bibitem{Wannier}  
G.H. Wannier, Phys. Rev. {\bf 52,} 191 (1937).  
  
  
\bibitem{Uffink}  
J. Hilgevoord and J.B.M. Uffink,  
    Phys. Lett. A {\bf 95,} 474 (1983);   
J.B.M. Uffink and J. Hilgevoord, ibid {\bf 105,} 176 (1984);  
Found. Phys. {\bf 15,}  925 (1985).   
  
  
\bibitem{Greiner}  
M. Greiner et al., Nature {\bf 415,} 39 (2002).  
  
\bibitem{comput} 
S. Lloyd and J.-J.E.  Slotine, Phys. Rev. Lett. {\bf 80,} 4088 (1998), S.L. 
Braunstein, Nature {\bf 394,} 47 (1998); S. Lloyd and S.L. Braunstein, Phys. 
Rev. Lett. {\bf 82,} 1784 (1999). 
  
\end{references}
\end{document}